\documentclass{article}
\usepackage{arxiv}
\pdfoutput=1

\usepackage{amssymb}
\usepackage{mathtools}

\usepackage{multirow}

\usepackage[dvipsnames, table]{xcolor}

\usepackage[shortlabels]{enumitem}
\usepackage{hyperref}
\usepackage{setspace}
\usepackage{comment}
\usepackage[most]{tcolorbox}

\renewcommand{\min}[1]{\mathchoice
  {\underset{#1}{\operatorname{min}}\,}%
  {\operatorname{min}_{#1}}%
  {\operatorname{min}_{#1}}%
  {\operatorname{min}_{#1}}%
}
\renewcommand{\max}[1]{\mathchoice
  {\underset{#1}{\operatorname{max}}\,}%
  {\operatorname{max}_{#1}}%
  {\operatorname{max}_{#1}}%
  {\operatorname{max}_{#1}}%
}
\newcommand{\argmin}[1]{\mathchoice
  {\operatorname{arg}\, \min{#1}}%
  {\operatorname{arg}\, \min{#1}}%
  {\operatorname{arg}\, \min{#1}}%
  {\operatorname{arg}\, \min{#1}}%
}

\renewcommand{\exp}[1]{e^{#1}}

\newcommand{\eqdef  }{\triangleq}
\newcommand{\rr}{\mathbb{R}}

\newcommand{\simiid}{\overset{\mathrm{i.i.d.}}{\sim}}
\newcommand{\diag}[1]{\operatorname{diag}\!\left\{#1\right\}}

\newcommand{\quotes}[1]{``{#1}''}
\newcommand{\citeauthor}[2]{#1 et al. \cite{#2}}

\newcommand{\Liion}{Li-ion} %
\newcommand{\BMS}{BMS} %
\newcommand{\EV}{EV} %
\newcommand{\SOC}{{SOC}} %
\newcommand{\KF}{KF} %
\newcommand{\EKF}{E\KF{}} %
\newcommand{\AEKF}{A\EKF{}} %
\newcommand{\ECM}{ECM} %
\newcommand{\LPV}{LPV} %
\newcommand{\APV}{APV} %
\newcommand{\ARX}{ARX} %
\newcommand{\BBO}{BBO} %
\newcommand{\ANN}{ANN} %
\newcommand{\UKF}{U\KF{}} %
\newcommand{\OCV}{OCV} %
\newcommand{\GEIS}{GEIS} %
\newcommand{\MMAE}{MMAE} %
\newcommand{\LTI}{LTI} %
\newcommand{\RMSE}{RMSE} %
\newcommand{\TV}{TV} %
\newcommand{\FFNN}{FFNN} %
\newcommand{\BEKF}{B\EKF{}} %
\newcommand{\VS}{VS} %
\newcommand{\VSF}{VSF} %
\newcommand{\settablestretch}{\setstretch{1.2}}
\newcommand{\settablefontsize}{\footnotesize}
\newcommand{\colorrowoftable}{\rowcolor{gray!40}}

\allowdisplaybreaks[1]

\begin{document}

\onecolumn

\title{A virtual sensor fusion approach for state of charge estimation of lithium-ion cells}

\author{
  Davide Previtali\\
  Department of Management, Information and Production Engineering\\
  University of Bergamo (Via G. Marconi 5, 24044, Dalmine (BG), Italy)\\
  \texttt{davide.previtali@unibg.it}\\
  \And
  Daniele Masti\\
  Gran Sasso Science Institute (Viale Francesco Crispi, 7, 67100 L'Aquila (AQ), Italy)\\
  \texttt{daniele.masti@gssi.it}\\
  \And
  Mirko Mazzoleni\\
  Department of Management, Information and Production Engineering\\
  University of Bergamo (Via G. Marconi 5, 24044, Dalmine (BG), Italy)\\
  \texttt{mirko.mazzoleni@unibg.it}\\
  \And
  Fabio Previdi\\
  Department of Management, Information and Production Engineering\\
  University of Bergamo (Via G. Marconi 5, 24044, Dalmine (BG), Italy)\\
  \texttt{fabio.previdi@unibg.it}\\
}

\date{}

\twocolumn[
  \maketitle

  \begin{abstract}
    This paper addresses the estimation of the State Of Charge (\SOC{}) of lithium-ion cells via the combination of two widely used paradigms: Kalman Filters (\KF{}s) equipped with Equivalent Circuit Models (\ECM{}s) and %
    machine-learning approaches. In particular, a recent Virtual Sensor (\VS{}) synthesis technique is considered, which operates as follows: (i) learn an Affine Parameter-Varying (\APV{}) model of the cell directly from data, (ii) derive a bank of linear observers from the \APV{} model, (iii) train a machine-learning technique from features extracted from the observers together with input and output data to predict the \SOC{}. The \SOC{} predictions returned by the \VS{} are supplied to an Extended \KF{} (\EKF{}) as output measurements along with the cell terminal voltage, %
    combining the two paradigms. A data-driven calibration strategy for the noise covariance matrices of the \EKF{} is proposed. Experimental results show that the designed %
    approach is beneficial w.r.t. \SOC{} estimation accuracy and smoothness.

    \keywords{
      Lithium-ion cell, State of charge estimation, Virtual sensor.
    }
  \end{abstract}

  \noindent\rule{8.4cm}{1pt}\\
  This is the authors' version of a paper accepted at \textit{IECON 2025 - 51st Annual Conference of the IEEE Industrial Electronics Society conference}. It is posted here for your personal use, not for redistribution.\\
  Please cite the conference version once available. \\
  D. Previtali, D. Masti, M. Mazzoleni and F. Previdi, \quotes{A virtual sensor fusion approach for state of charge estimation of lithium-ion cells} in \textit{IECON 2025 - 51st Annual Conference of the IEEE Industrial Electronics Society conference. IEEE, 2025}.
]

\clearpage
\newpage
You may use the following bibtex entry:
\begin{tcolorbox}
  @inproceedings\{previtali2025virtual, \\
  title = \{A virtual sensor fusion approach for state of charge estimation of lithium-ion cells\}, \\
  author = \{Previtali, Davide and Masti, Daniele and Mazzoleni, Mirko and Previdi, Fabio\}, \\
  booktitle =\{IECON 2025 - 51st Annual Conference of the IEEE Industrial Electronics Society conference\}, \\
  year = \{2025\}, \\
  organization = \{IEEE\},\\
  \}
\end{tcolorbox}
\section{Introduction}
\label{sec:introduction}
Lithium-ion (\Liion{}) battery technology is one of the most promising solutions for vehicle electrification %
due to its high energy density, low self-discharge rate, and long cycle life~\cite{goodenough2011challenges}. %
Electric Vehicles (\EV{}s) rely on large battery packs composed of hundreds of \Liion{} cells connected in series and/or in parallel to meet energy and power requirements~\cite{farmann2018comparative}. The safe operation of a battery pack is overseen by the so-called Battery Management System (\BMS{}), which combines hardware and software components for disparate purposes~\cite{plett2015battery_volII}. One of the core functionalities of \BMS{}s is State Of Charge (\SOC{}) estimation, indicating the amount of energy available in the battery and, consequently, the residual driving range of an \EV{}. Therefore, the design of \SOC{} estimation algorithms at the cell and battery pack level has received much attention in the past two decades. Focusing on cell-level state of charge monitoring, the most prominent algorithms are %
Kalman Filters (\KF{}s), constituting more than $50\%$ of the relevant literature~\cite[Figure 3]{shrivastava2019overview}. Particularly, given the nonlinear behavior of \Liion{} cells, the Extended Kalman Filter (\EKF{}) is one of the most employed \KF{} formulations. For example,%
~\citeauthor{Chen}{chen2012state} applied an \EKF{} based on an Equivalent-Circuit cell Model (\ECM{}) to estimate the \SOC{} of a pouch \Liion{} cell for plug-in hybrid \EV{}s.%
~\citeauthor{Yun}{yun2023state} enriched the just mentioned \ECM{} by considering its resistance and capacitance parameters as \SOC{}-dependent, improving state of charge estimation accuracy when applied in conjunction with the \EKF{}. Finally,%
~\citeauthor{Taborelli}{taborelli2014state} designed an \ECM{}-based Adaptive Extended Kalman Filter (\AEKF{}) for \SOC{} estimation purposes, showing better performance compared to the baseline \EKF{} due to its automatic adaptation of the noise covariance matrices.

Although Kalman-filter-based state of charge monitoring strategies are currently the most popular, %
machine-learning approaches, such as Artificial Neural Networks (\ANN{}s), are rapidly becoming attractive alternatives to \KF{}s due to their remarkable accuracy with negligible modeling effort. Nonetheless, in this context, %
\ANN{}s are rarely applied as is and they often require some correction mechanism to improve generalizability and \SOC{} estimation smoothness. For example,%
~\citeauthor{Liu}{liu2015improved} implemented a voltage correction strategy to improve \ANN{} performance. Instead,%
~\citeauthor{He}{he2014state} replaced the equivalent-circuit model with an \ANN{} estimated from data and combined the neural network with an Unscented Kalman Filter (\UKF{}) to achieve smoother \SOC{} estimates compared to the baseline \ANN{}.%
~\citeauthor{Masti}{masti2021machine} designed a machine-learning approach for virtual sensor synthesis of parameter-varying systems whose parameters depend on a set of scheduling variables (e.g., the \SOC{} in the case of \Liion{} cells). The method %
derives directly from data an Affine Parameter-Varying (\APV{}) AutoRegressive with eXogenous inputs (\ARX{}) model whose parameters depend on the scheduling variables. A bank of observers is designed from the \APV{}~\ARX{} model. Then, features are extracted from the observers %
and subsequently fed to an \ANN{} along with the inputs and outputs of the system to predict the scheduling variables. The method in~\cite{masti2021machine} was tested on simulated \Liion{} cell data, exhibiting better bandwidth and state of charge estimation accuracy compared to the \EKF{} but with a lower degree of smoothness.

In a fashion similar to~\cite{he2014state}, this paper %
combines a machine-learning approach with the Kalman filter paradigm. Specifically, we fuse the method in~\cite{masti2021machine} with an \EKF{} %
to improve state of charge estimation accuracy while maintaining a high degree of smoothness. Furthermore, we propose a novel data-driven calibration strategy for the noise covariance matrices of the Kalman filter based on Black-Box Optimization (\BBO{})~\cite[Chapter 2]{previtali2024phdthesis}. The proposed %
approach is validated on \Liion{} cell experimental data.

The rest of this paper is organized as follows. Section~\ref{sec:experimental_setup} describes the lithium-ion cell under study and available experimental data. Section~\ref{sec:ECM} presents the equivalent-circuit model employed by the \EKF{}. Section~\ref{sec:data_driven_approach} reviews the virtual sensor synthesis method in~\cite{masti2021machine}. Section~\ref{sec:sensor_fusion_approach} presents the proposed virtual sensor fusion approach; its \SOC{} estimation performance is analyzed in Section~\ref{sec:experimental_results}. Finally, Section~\ref{sec:conclusion} is devoted to concluding remarks.

\section{Experimental setup}
\label{sec:experimental_setup}
In this work, we consider the Samsung INR21700-50E cylindrical lithium-ion cell with technical specifications reported in \tablename{}~\ref{tab:cell_specifications}. Experiments are performed using a BioLogic VSP-3e potentiostat connected to a BioLogic FlexP0060 booster to achieve the current ranges that the cell under study can sustain. The BioLogic VSP-3e potentiostat is controlled by a PC with EC-Lab software via an Ethernet cable. 
Experiments are carried out in a temperature-controlled environment with temperatures ranging between $21^\circ \mathrm{C}$ and $25^\circ \mathrm{C}$. The sampling time of the equipment is $\tau_{\mathrm{s}} = 1\,\mathrm{s}$.
\begin{table}[!htb]
    \centering
    \caption{Technical specifications of the Samsung INR21700-50E lithium-ion cell.}
    \settablestretch
    \settablefontsize
    \label{tab:cell_specifications}

    \begin{tabular}{cc}
        \textbf{Parameter}                                 & \textbf{Value}\tabularnewline
        \hline
        \colorrowoftable Nominal voltage                   & $3.6\,\mathrm{V}$\tabularnewline
        Nominal discharge capacity                         & $4.9\,\text{A} \mathrm{h}$\tabularnewline
        \colorrowoftable Charge cut-off voltage $v_{\max{}}$            & $4.2\,\mathrm{V}$ (at $23^{\circ}\mathrm{C}$)\tabularnewline
        Discharge cut-off voltage $v_{\min{}}$                        & $2.5\,\mathrm{V}$ (at $23^{\circ}\mathrm{C}$)\tabularnewline
        \colorrowoftable Maximum continuous charge current & $4.9\,\text{A}$ ($1\mathrm{C}$)\tabularnewline
        Maximum continuous discharge current               & $9.8\,\text{A}$ ($2\mathrm{C}$)\tabularnewline
        \hline
    \end{tabular}
\end{table}

Notation-wise, in what follows, we denote the discrete-time index as $k \in \mathbb{N}$\footnote{We consider $0 \in \mathbb{N}$.}, the cell terminal voltage at the time $k \tau_{\mathrm{s}}$ as $v[k] \in \mathbb{R}_{\geq 0}$ (in $\mathrm{V}$), and the load current as $i[k] \in \mathbb{R}$ (in $\mathrm{A}$), $i[k]> 0$ during discharging and $i[k] < 0$ during charging. The acquired data is enriched with the state of charge signal, denoted as $\mathrm{\SOC{}}[k] \in \left[0, 1\right]$, obtained via the Coulomb counting method~\cite[Chapter 1]{plett2015battery_volII}.

To estimate the parameters of equivalent-circuit models for Kalman filter purposes (Section \ref{sec:ECM}), two experiments are conducted: a low-current Open Circuit Voltage (\OCV{}) trial~\cite[Chapter 2]{plett2015battery_volI} and Galvanostatic Electrochemical Impedance Spectroscopy (\GEIS{})~\cite{lazanas2023electrochemical}. Instead, several dynamic current profiles experiments~\cite{he2014state} are carried out to calibrate the Kalman filters, train the machine-learning approach, and assess the accuracy of the methods under study (Sections \ref{sec:data_driven_approach}, \ref{sec:sensor_fusion_approach}, \ref{sec:experimental_results}). Before each trial, the \Liion{} cell is fully charged %
following the protocol described in its datasheet.

\textit{Low-current \OCV{} experiment.}
The fully-charged cell is first discharged at low $\mathrm{C}$-rate, namely $\frac{\mathrm{C}}{20}$ ($250\,\mathrm{mA}$), until reaching the discharge cut-off voltage $v_{\min{}}$ in \tablename{}~\ref{tab:cell_specifications}. Then, after a resting period of $3\,\mathrm{h}$, the cell is charged at the same $\mathrm{C}$-rate until reaching the manufacturer-specified charge cut-off voltage $v_{\max{}}$. We denote the datasets obtained from this experiment as
\begin{subequations}
    \label{eq:LC_OCV_data}
    \begin{align}
        &\!\!\!\!\!\mathcal{D}_{\mathrm{d}} \! = \! \left\{\left(v[k], i[k], \mathrm{\SOC{}}[k]\right) \!: \! k \! \in \! \{0, \ldots, N_{\mathrm{d}}\!-\!1\}, i[k] \! > \! 0 \right\}\!, \\
        &\!\!\!\!\!\mathcal{D}_{\mathrm{c}} \! = \! \left\{\left(v[k], i[k], \mathrm{\SOC{}}[k]\right) \! : \! k \! \in \! \{0, \ldots, N_{\mathrm{c}} \!- \!1\}, i[k] \! < \! 0 \right\}\!,
    \end{align}
\end{subequations}
for the discharging ($N_{\mathrm{d}} \in \mathbb{N}$ data in total) and charging ($N_{\mathrm{c}} \in \mathbb{N}$ data in total) portions respectively.

\textit{\GEIS{} experiment.}
Starting from a fully-charged cell, we discharge (roughly) $10\%$ of the state of charge at a $\mathrm{C}$-rate of $1\mathrm{C}$ ($4.9\,\mathrm{A}$). %
Afterwards, we let the cell rest for $3\,\mathrm{h}$ to let it reach the equilibrium. Then, we carry out a frequency sweep (\GEIS{}) from $10\,\mathrm{kHz}$ to $10\,\mathrm{mHz}$ via zero-mean sine waves with an amplitude of $\frac{\mathrm{C}}{5}$ ($1\,\mathrm{A}$). Ten frequencies per decade are considered, attaining a total of $60$ impedance spectrum points. %
The process of discharging the cell, letting it rest, and performing the \GEIS{} is repeated until reaching the %
discharge cut-off voltage $v_{\min{}}$. Let $\mathcal{S}$ denote the set of tested \SOC{} equilibria around which the frequency sweeps are carried out. We group the frequency-domain data attained at each $\bar{\mathrm{\SOC{}}} \in \mathcal{S}$ inside the datasets
\begin{equation}
    \label{eq:GEIS_data}
    \!\!\!\! \mathcal{D}_{\mathrm{GEIS}} \! \left(\bar{\mathrm{\SOC{}}}\right) \!\! = \!\!\left\{\!\left(\omega_n, \varsigma_n(\bar{\mathrm{\SOC{}}})\right) \! : \! n \! \in \!\mathbb{N}, n \! \in \! \{1, \ldots, N_{\mathrm{f}}\}\!\right\}\!,
\end{equation}
where $N_{\mathrm{f}} \in \mathbb{N}$ is the number of tested frequencies, $\omega_n \in \mathbb{R}_{>0}$ (in $\frac{\mathrm{rad}}{\mathrm{s}}$) being the $n$-th one, and $\varsigma_n(\bar{\mathrm{\SOC{}}}) \in \mathbb{C}$ (in $\Omega$) is the measured impedance at the frequency $\omega_n$ and for the equilibrium $\bar{\mathrm{\SOC{}}} \in \mathcal{S}$.

\begin{figure}[!htb]
	\centering
	\includegraphics[width=\columnwidth]{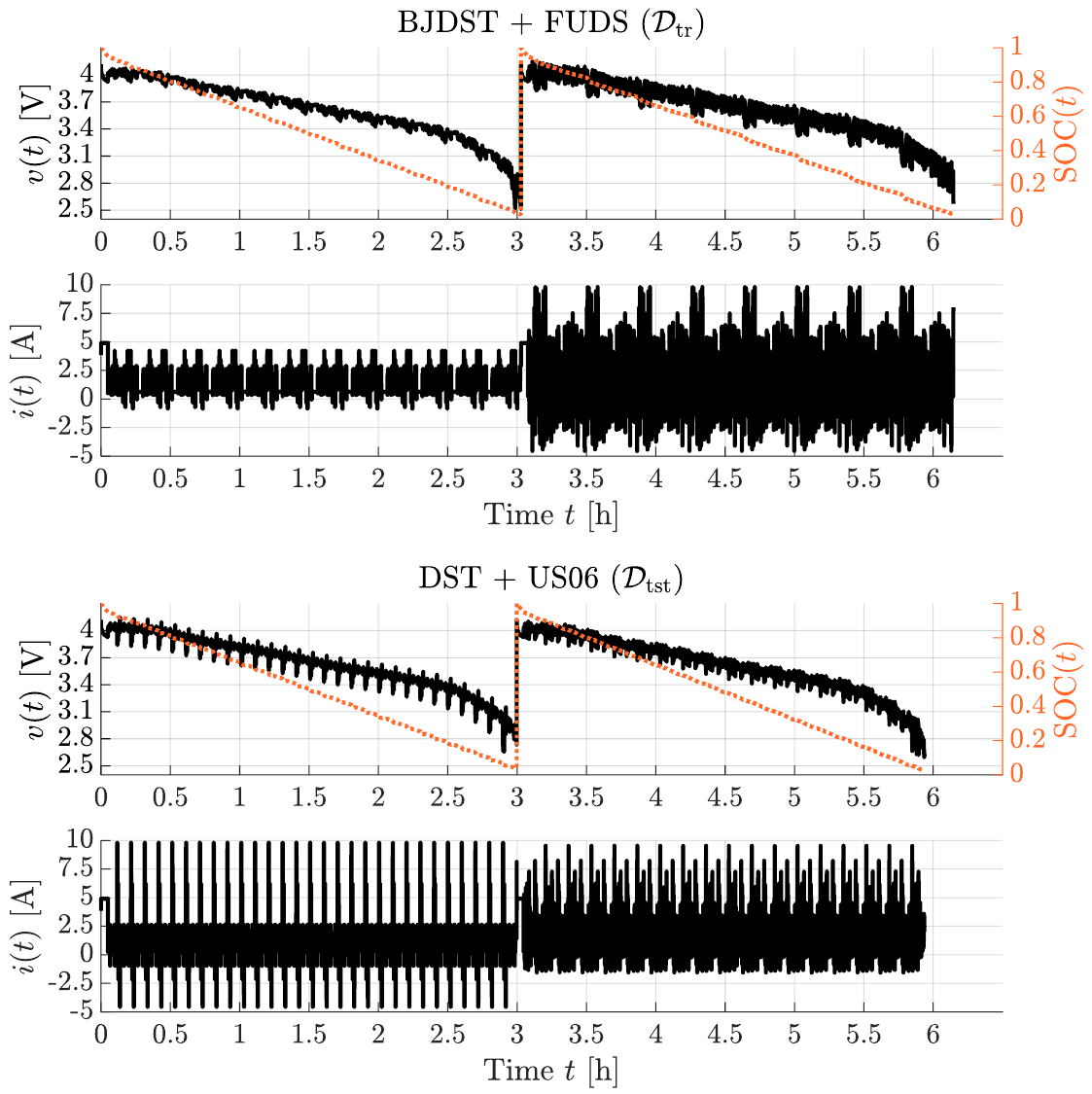}
	\caption{\label{fig:dynamic_current_profiles} Considered merged dynamic current profiles.}
\end{figure}

\textit{Dynamic current profiles.}
We apply four dynamic current profiles from the literature to the cell under study, i.e. the BJDST, DST, FUDS and US06 profiles~\cite{he2014state}. Before the application of each profile, the fully-charged cell is discharged by $5\%$ at a $\mathrm{C}$-rate of $1\mathrm{C}$ ($4.9\,\mathrm{A}$) to prevent overcharge. The experiment is stopped once the discharge cut-off voltage $v_{\min{}}$ in \tablename{}~\ref{tab:cell_specifications} is reached. For method training, calibration, and testing purposes, we merge %
the profiles in pairs, specifically the BJDST and FUDS ($N_{\mathrm{tr}} \in \mathbb{N}$ data in total), and the FUDS and US06 ($N_{\mathrm{tst}} \in \mathbb{N}$ data in total), obtaining the datasets
\begin{subequations}
    \label{eq:dynamic_current_profiles_data}
    \begin{align}
        \label{eq:training_dataset}
        \mathcal{D}_{\mathrm{tr}} \! &= \! \left\{\left(v[k], i[k], \mathrm{\SOC{}}[k]\right) \!: \! k \! \in \! \{0, \ldots, N_{\mathrm{tr}}-1\}\right\}, \\
        \label{eq:test_dataset}
        \mathcal{D}_{\mathrm{tst}} \! &= \! \left\{\left(v[k], i[k], \mathrm{\SOC{}}[k]\right) \! : \! k \! \in \! \{0, \ldots, N_{\mathrm{tst}}-1\}\right\},
    \end{align}
\end{subequations}
respectively, as depicted in \figurename{}~\ref{fig:dynamic_current_profiles}. By merging %
two profiles, we assess if the \SOC{} estimators are able to handle abrupt transitions (such as the ones that happen at $t \approx 3\, \mathrm{h}$, i.e. when switching from a profile to the other) that may arise if data are lost during operation, effectively making state of charge estimation more challenging.

\section{Equivalent-circuit cell model}
\label{sec:ECM}
Kalman filters require a discrete-time state-space model for the \Liion{} cell. Consistently with the reviewed literature, we employ an equivalent-circuit model, achieving a fair trade-off between accuracy and model complexity~\cite{shrivastava2019overview}. 
The \ECM{} considered in this work is the Thevenin model, which consists of an ideal \SOC{}-dependent voltage source $\mathrm{\OCV{}}\left(\mathrm{\SOC{}}\right) \in \mathbb{R}_{\geq 0}$ (in $\mathrm{V}$) representing the terminal voltage when the cell is at rest and no load is applied to it (i.e. the so-called open-circuit voltage), followed by a \SOC{}-dependent series resistance $R_0\left(\mathrm{\SOC{}}\right) \in \mathbb{R}_{> 0}$ (in $\Omega$), and one $RC$ network with \SOC{}-dependent resistance $R_1\left(\mathrm{\SOC{}}\right) \in \mathbb{R}_{> 0}$ (in $\Omega$) and capacitance $C_1\left(\mathrm{\SOC{}}\right) \in \mathbb{R}_{> 0}$ (in $\mathrm{F}$). 
Let $i_{R_1} \in \mathbb{R}$ (in $\mathrm{A}$) be the current flowing in $R_1\left(\mathrm{\SOC{}}\right)$. Then, we can write down the Thevenin model in state-space form as~\cite[Chapter 2]{plett2015battery_volI}:
\begin{equation}
    \label{eq:ECM_discrete_time}
    \!\!\!\!\!\!\!\!\begin{cases}
        \mathrm{\SOC{}}\! \left[k\!+\!1\right]\!=\! \mathrm{\SOC{}}\left[k\right]\!-\!\frac{\tau_{\mathrm{s}}}{Q} \eta\left[k\right] i\left[k\right]\!,                                                 \\
        {i}_{R_{1}} \! \left[k\!+\!1\right]\!=\!\exp{-\frac{\tau_{\mathrm{s}}}{\tau_1\left(\mathrm{\SOC{}}[k]\right)}} i_{R_{1}} \! \left[k\right] \! + \! \left(\!1 \! - \! \exp{-\frac{\tau_{\mathrm{s}}}{\tau_1\left(\mathrm{\SOC{}}[k]\right)}}\!\right) \! i\left[k\right]\!,                                               \\
        v\left[k\right] \! = \! \mathrm{\OCV{}} \left(\mathrm{\SOC{}}\left[k\right]\right) \! - \! R_{1}\left(\mathrm{\SOC{}}\left[k\right]\right)  i_{R_{1}}\left[k\right] \! + \!\\
        \qquad \ - R_{0}\left(\mathrm{\SOC{}}\left[k\right]\right) i\left[k\right]\!,
    \end{cases}
\end{equation}
where $Q \in \mathbb{R}_{\geq 0}$ (in $\mathrm{A} \mathrm{s}$) is the cell total capacity, i.e. the total amount of charge that the cell can store, $\eta[k] \in [0, 1]$ is the Coulombic efficiency, which is such that $\eta[k] = 1$ when $i[k] \geq 0$ (during discharging) and $\eta[k] = \eta_{\mathrm{c}}$, $\eta_{\mathrm{c}} \in [0, 1]$, when $i[k] < 0$ (during charging), and $\tau_1\left(\mathrm{\SOC{}}\right) = R_1\left(\mathrm{\SOC{}}\right) C_1\left(\mathrm{\SOC{}}\right)$ (in $\mathrm{s}$) is the time constant of the $RC$ network.

\begin{figure}[!htb]
	\centering
	\includegraphics[width=\columnwidth]{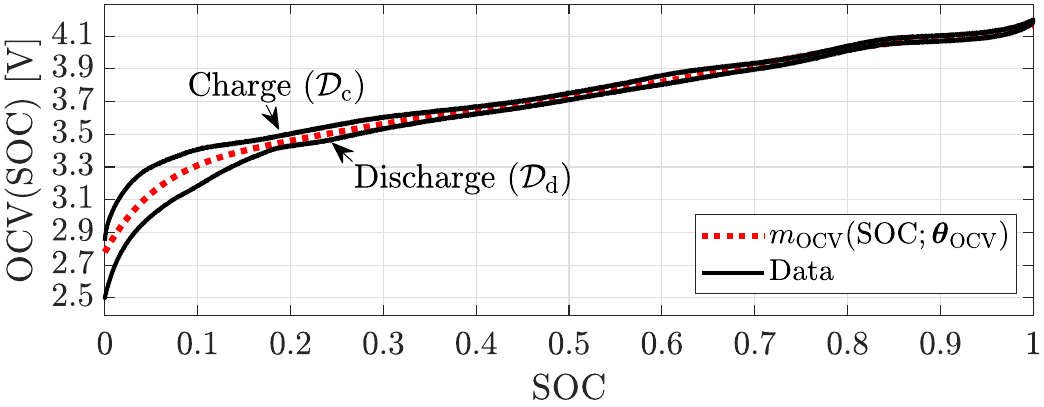}
	\caption{\label{fig:OCV_SOC_curve} Estimated $\mathrm{\OCV{}} \left(\mathrm{\SOC{}}\right)$ curve.}
\end{figure}

\textit{Identification.}
The details behind the estimation of the parameters of the \ECM{} in~\eqref{eq:ECM_discrete_time} are out of scope of this paper; the interested reader is referred to~\cite[Chapter 2]{plett2015battery_volI},~\cite{lazanas2023electrochemical}, and~\cite{santoni2024guide}. In short, using the low current \OCV{} experiment data (Section~\ref{sec:experimental_setup}), the cell total capacity $Q$ is estimated by integrating the current measurements in $\mathcal{D}_{\mathrm{d}}$ in~\eqref{eq:LC_OCV_data}, while $\eta_{\mathrm{c}}$ is simply the ratio between the total amount of ampere-hours discharged in $\mathcal{D}_{\mathrm{d}}$ (i.e., $Q$) and the total amount of ampere-hours charged in $\mathcal{D}_{\mathrm{c}}$. %
The $\mathrm{\OCV{}} \left(\mathrm{\SOC{}}\right)$ curve in~\eqref{eq:ECM_discrete_time} is approximated as a polynomial with degree $n_{\mathrm{OCV}} \in \mathbb{N}$ and coefficients 
$\boldsymbol{\theta}_{\mathrm{\OCV{}}} = \left[\theta_{0, \mathrm{\OCV{}}}, \ldots, \theta_{n_{\mathrm{\OCV{}}}, \mathrm{\OCV{}}} \right]^\top \in \mathbb{R}^{n_{\mathrm{\OCV{}}} + 1}$:
\begin{equation}
    \label{eq:OCV_model}
    m_{\mathrm{\OCV{}}}\left(\mathrm{\SOC{}}; \boldsymbol{\theta}_{\mathrm{\OCV{}}}\right) = \sum_{n = 0}^{n_{\mathrm{\OCV{}}}} \theta_{n, \mathrm{\OCV{}}} \mathrm{\SOC{}}^n.
\end{equation}
The coefficients $\boldsymbol{\theta}_{\mathrm{\OCV{}}}$ are estimated by minimizing the difference between $m_{\mathrm{\OCV{}}}\left(\mathrm{\SOC{}}; \boldsymbol{\theta}_{\mathrm{\OCV{}}}\right)$ and the data in $\mathcal{D}_{\mathrm{d}} \cup \mathcal{D}_{\mathrm{c}}$ in a least squares sense. In our case, we have chosen $n_{\mathrm{\OCV{}}} = 8$, leading to the results in \figurename{}~\ref{fig:OCV_SOC_curve}.

\begin{figure*}[!htb]
	\centering
	\includegraphics[width=\textwidth]{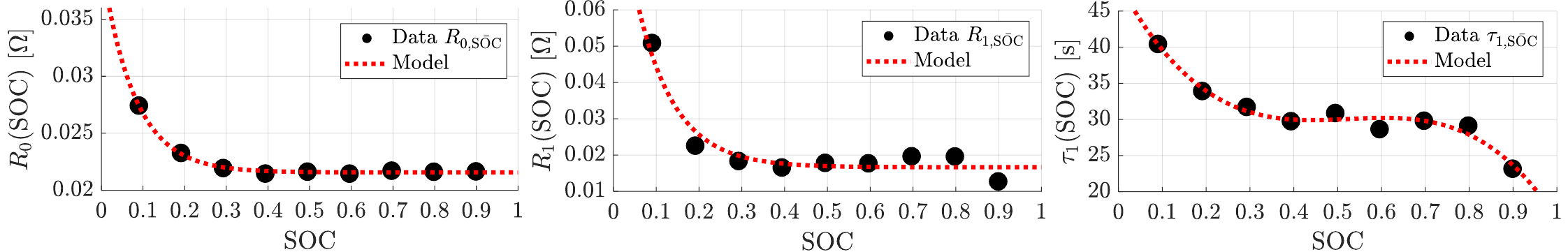}
	\caption{\label{fig:ECM_parameters} Estimated resistances and time constant curves for the considered equivalent-circuit model.}
\end{figure*}

Instead, the resistances $R_0\left(\mathrm{\SOC{}}\right)$, $R_1\left(\mathrm{\SOC{}}\right)$, and time constant $\tau_1\left(\mathrm{\SOC{}}\right)$ for the \ECM{} in~\eqref{eq:ECM_discrete_time} are estimated from \GEIS{} data (Section~\ref{sec:experimental_setup}). In particular, at each equilibria $\bar{\mathrm{\SOC{}}} \in \mathcal{S}$ and in the Laplace domain, the \ECM{} in~\eqref{eq:ECM_discrete_time} can be linearized as~\cite{santoni2024guide}:
\begin{equation}
    \label{eq:linearized_ECM}
    G\left(s; \boldsymbol{\theta}_{\bar{\mathrm{\SOC{}}}}, \bar{\mathrm{\SOC{}}}\right) = R_{0, \bar{\mathrm{\SOC{}}}} + \frac{R_{1, \bar{\mathrm{\SOC{}}}}}{\tau_{1, \bar{\mathrm{\SOC{}}}} s + 1},
\end{equation}
with %
$\boldsymbol{\theta}_{\bar{\mathrm{\SOC{}}}} = \left[ R_{0, \bar{\mathrm{\SOC{}}}}, R_{1, \bar{\mathrm{\SOC{}}}}, \tau_{1, \bar{\mathrm{\SOC{}}}} \right]^\top \in \mathbb{R}_{>0}^3$ being the parameters of interest at the equilibrium $\bar{\mathrm{\SOC{}}}$. Then, we estimate $\boldsymbol{\theta}_{\bar{\mathrm{\SOC{}}}}$ at each $\bar{\mathrm{\SOC{}}} \in \mathcal{S}$ by minimizing the difference between the frequency response of~\eqref{eq:linearized_ECM} evaluated at $\omega_n, n \in \{1, \ldots, N_{\mathrm{f}}\}$, and the data $\mathcal{D}_{\mathrm{GEIS}}\left(\bar{\mathrm{\SOC{}}}\right)$ in~\eqref{eq:GEIS_data} in a least squares sense\footnote{Only data with non-positive imaginary part are considered.}~\cite{santoni2024guide}. Afterwards, %
we fit the following exponential and polynomial curves:
\begin{subequations}
    \label{eq:R_tau_parameters_models}
    \begin{align}
        R_j(\mathrm{\SOC{}}; \boldsymbol{\theta}_{R_j}) \! &= \! \theta_{1, R_j}  \exp{- \theta_{2, R_j} \mathrm{\SOC{}}} \! + \! \theta_{3, R_j}, j \in \{0, 1\},\\
        \tau_1(\mathrm{\SOC{}}; \boldsymbol{\theta}_{\tau_1}) \! &= \! \sum_{n = 0}^{n_{\tau}} \theta_{n, \tau_1} \mathrm{\SOC{}}^n,
    \end{align}
\end{subequations}
where %
$\boldsymbol{\theta}_{R_j} = \left[\theta_{1, R_j}, \theta_{2, R_j}, \theta_{3, R_j} \right]^\top \in \mathbb{R}^3$ and %
$\boldsymbol{\theta}_{\tau_1} = \left[ \theta_{0, \tau_1}, \, \theta_{n_{\tau}, \tau_1}\right]^\top \in \mathbb{R}^{n_{\tau} + 1}, n_{\tau} = 3,$ are coefficients estimated by minimizing the difference (in a least squares sense) between the curves in~\eqref{eq:R_tau_parameters_models} and the values of $\boldsymbol{\theta}_{\bar{\mathrm{\SOC{}}}}$ estimated from \GEIS{} data. The results are shown in \figurename{}~\ref{fig:ECM_parameters}.

\section{Data-driven virtual sensor synthesis}
\label{sec:data_driven_approach}
As we have seen in Section~\ref{sec:ECM}, developing a phenomenological cell model such as~\eqref{eq:ECM_discrete_time} to be employed in \KF{} schemes for state of charge estimation is a formidable problem, requiring a mixture of data coming from ad hoc time-consuming experiments and data fitting techniques. Nonetheless, the \ECM{} in~\eqref{eq:ECM_discrete_time} is only an approximation of true \Liion{} cell behavior, which involves complex microscale dynamics related to charge and mass conservation as well as lithium diffusion mechanisms~\cite[Chapter 3]{plett2015battery_volI}. A more general cell model is the following:
\begin{equation}
  \label{eq:general_cell_model}
  \mathcal{M}_\mathrm{cell} \eqdef \begin{cases}
    \mathrm{\SOC{}}\left[k+1\right] = h(\mathrm{\SOC{}}\left[k\right], i\left[k\right]),                                                 \\
    \boldsymbol{x}\left[k+1\right] = \boldsymbol{f}(\boldsymbol{x}\left[k\right], i\left[k\right], \mathrm{\SOC{}}\left[k\right]), \\
    v\left[k\right] = g(\boldsymbol{x}\left[k\right], i\left[k\right], \mathrm{\SOC{}}\left[k\right]),
\end{cases}
\end{equation}
where $\boldsymbol{x}[k] \in \mathbb{R}^{n_x}$, $n_x \in \mathbb{N}$, is a state vector encompassing all \Liion{} cell dynamics of interest (excluding the \SOC{}), and $h: [0, 1] \times \mathbb{R} \to [0, 1]$, $\boldsymbol{f}: \mathbb{R}^{n_x} \times \mathbb{R} \times [0, 1] \to \mathbb{R}^{n_x}$, $g: \mathbb{R}^{n_x} \times \mathbb{R} \times [0, 1] \to \mathbb{R}$, are the state (\SOC{} and $\boldsymbol{x}$) and output equations respectively. However, the definition of $h, \boldsymbol{f},$ and $g$ in~\eqref{eq:general_cell_model} is fairly complex, especially when derived from first principles~\cite{plett2015battery_volI}.

Virtual sensors%
~\cite{milanese2009filter} provide an alternative approach to solving the \SOC{} estimation problem: the idea is to forego the definition of a cell model such as~\eqref{eq:ECM_discrete_time} or~\eqref{eq:general_cell_model} for \KF{} schemes and instead build an end-to-end \SOC{} estimator by learning directly from data.
In the following, we exploit such a venue and build from the proposal in~\cite{masti2021machine,masti2019learning} to develop a virtual sensor inspired by the so-called Multiple Model Adaptive Estimation (\MMAE{}~\cite{akca2019multipleModelSurvey}) paradigm. Doing so requires a three-step procedure:
\begin{enumerate}[(\texttt{S}1)]
  \item \label{item:MMAE_1} Learn a finite set of simple Linear Time-Invariant (\LTI{}) models from data that roughly covers the electrical behavior of the cell for the entire \SOC{} range;
  \item \label{item:MMAE_2} Design a set of standard linear observers based on such models;
  \item \label{item:MMAE_3} Use a machine-learning method to predict the \SOC{} from the estimates obtained by the observers plus the raw cell input/output signals.
\end{enumerate}
The intuition behind this approach is that for many systems including the considered scenario, the dynamics of some states (e.g., $\boldsymbol{x}$ in~\eqref{eq:general_cell_model}) %
are much faster than the ones of other states (e.g., $\mathrm{\SOC{}}$ in~\eqref{eq:general_cell_model}) up to the point in which the dynamics of the latter can be neglected and considered as %
scheduling variables of the system. For these kinds of systems, multiple model filtering has proved itself extremely successful, yet its classical formulation requires strong statistical prior knowledge and good knowledge of the underlying system dynamics. Instead,~\citeauthor{Masti}{masti2021machine} proposed to learn such knowledge directly from data. %
To do so, as a first step we need to reconcile the \MMAE{} framework with the nonlinear model $\mathcal{M}_\mathrm{cell}$ in~\eqref{eq:general_cell_model}. 

For that purpose, assume that $\boldsymbol{f}$ and $g$ are differentiable. Then, in the neighborhood of an arbitrary tuple $(\bar{\boldsymbol{x}}, \bar{i}, \bar{\mathrm{SOC}})$ it is possible to approximate $\mathcal{M}_\mathrm{cell}$ in~\eqref{eq:general_cell_model} as:
\begin{subequations}
  \label{eq:linearized-model}
  \begin{align}
    \boldsymbol{x}[k\!+\!1] & \! \approx \!\boldsymbol{f}(\bar{\boldsymbol{x}}, \bar{i}, \bar{\mathrm{SOC}}) \! + \! \nabla_{\boldsymbol{x}} \boldsymbol{f}(\bar{\boldsymbol{x}}, \bar{i}, \bar{\mathrm{SOC}}) (\boldsymbol{x}[k] \! - \! \bar{\boldsymbol{x}}) + \\
                            & \quad + \nabla_i \boldsymbol{f}(\bar{\boldsymbol{x}}, \bar{i}, \bar{\mathrm{\SOC{}}}) (i[k] \! - \! \bar{i}),         \notag                                                                                       \\
    v[k]                    & \! \approx \!g(\bar{\boldsymbol{x}}, \bar{i}, \bar{\mathrm{\SOC{}}}) \! + \! \nabla_{\boldsymbol{x}} g(\bar x,\bar i,\bar{\SOC}) (\boldsymbol{x}[k] \! - \! \bar{\boldsymbol{x}}) +                       \\
                            & \quad + \nabla_i g(\bar{\boldsymbol{x}}, \bar{i}, \bar{\mathrm{SOC}}) (i[k] \! - \! \bar{i}). \notag
  \end{align}
\end{subequations}
In~\eqref{eq:linearized-model}, the contributions of the Jacobians w.r.t. \SOC{} are neglected as we assume this quantity to move slowly enough to remain close to $\bar{\mathrm{\SOC{}}}$ within a certain time interval, i.e. $\mathrm{\SOC{}}[k]\!-\!\bar{\mathrm{\SOC{}}}\!\approx \!0$. Hence, from~\eqref{eq:linearized-model}, we can derive the following affine parameter-varying approximation of~\eqref{eq:general_cell_model}:
\begin{subequations}
  \label{eq:APV}
  \begin{align}
    &\!\!\!\!\!\boldsymbol{x}[k\!+\!1] \!\! \approx \!\! A(\mathrm{\SOC{}}[k]) \boldsymbol{x}[k] \!\! + \!\! B(\mathrm{\SOC{}}[k]) i[k] \! + \! d(\mathrm{\SOC{}}[k]), \\
    &v[k] \! \approx \! C(\mathrm{\SOC{}}[k]) \boldsymbol{x}[k] \! + \! D(\mathrm{\SOC{}}[k]) i[k] \! + \! e(\mathrm{\SOC{}}[k]),
  \end{align}
\end{subequations}

If $\mathcal{M}_\mathrm{cell}$ in~\eqref{eq:general_cell_model} were known, a \MMAE{} scheme could be used to compute the likelihood that the system is operating around a tuple $(\bar{\boldsymbol{x}}, \bar{i}, \bar{\mathrm{\SOC{}}}_j)$, with $\bar{\mathrm{\SOC{}}}_j \in \Theta^{\mathrm{\SOC{}}} \eqdef  \{\bar{\mathrm{\SOC{}}}_1, \dots,\bar{\mathrm{\SOC{}}}_{N_{\theta}}\}$, $N_{\theta} \in \mathbb{N}$. %
However, this knowledge is not available. For this reason, in the following we opt to learn the linear behavior of $\mathcal{M}_\mathrm{cell}$ around a given \SOC{} directly from data.

\textit{Learning the local models~\ref{item:MMAE_1}.}
As in~\cite{masti2021machine}, %
we restrict our analysis to learning affine \ARX{} models of fixed order $M \in \mathbb{N}$, each of them uniquely identified by a parameter vector $\boldsymbol{\gamma} \in \rr^{n_\gamma}$, %
$n_\gamma = 2 M + 1$. Learning an \APV{} approximation of $\mathcal{M}_\mathrm{cell}$ in~\eqref{eq:general_cell_model} amounts to training a functional approximator $M_{\mathrm{\LPV{}}}:\rr\to\rr^{n_\gamma}$ to predict the correct vector $\boldsymbol{\gamma}_j \in \rr^{n_\gamma}$ corresponding to any given $\bar{\mathrm{\SOC{}}}_j \in \Theta^{\mathrm{\SOC{}}}$. To that end, given a training dataset $\mathcal{D}_{\mathrm{tr}}$ such as the one in~\eqref{eq:training_dataset}, $M_{\mathrm{\LPV{}}}$ is learnt %
by solving: %

\begin{align}
  \label{eq:MLPV_learning}
  &\min{M_{\mathrm{\LPV{}}}} \sum_{k = M}^{N_{\mathrm{tr}} - 1} \mathcal{L}_{M_{\mathrm{\LPV{}}}}(\hat{v}[k], v[k]) \\
  \mathrm{s.t.}\ & \hat{v}[k] \! = \! \boldsymbol{\varphi}[k]^\top \boldsymbol{\gamma}[k] \nonumber \\
  & \boldsymbol{\varphi}[k] \! = \! \left[-v[k\!-\!M], \ldots, -v[k\!-\!1], i[k\!-\!M], \ldots, i[k\!-\!1], 1\right]^\top \nonumber \\
  & \boldsymbol{\gamma}[k] \! = \! M_{\mathrm{\LPV{}}}(\mathrm{\SOC{}}[k]) \nonumber \\
  & k \in \{M, \ldots, N_{\mathrm{tr}} - 1\}, \nonumber
\end{align}
where $\mathcal{L}_{M_{\mathrm{\LPV{}}}}: \mathbb{R} \times \mathbb{R} \to \mathbb{R}$ is an appropriate loss function. 
From a practical perspective, \ANN{}s prove to be a good choice for $M_{\mathrm{\LPV{}}}$ while $\mathcal{L}_{M_{\mathrm{\LPV{}}}}$ can simply be set as the squared error, i.e. $ \mathcal{L}_{M_{\mathrm{\LPV{}}}}(\hat{v}[k], v[k]) = (\hat{v}[k] - v[k])^2$, or the absolute error, i.e. $ \mathcal{L}_{M_{\mathrm{\LPV{}}}}(\hat{v}[k], v[k]) = |\hat{v}[k] - v[k]|$.%

\textit{Selecting the representative models~\ref{item:MMAE_1}.}
After solving~\eqref{eq:MLPV_learning}, a set $\Gamma \eqdef \{\boldsymbol{\gamma}[k]: k \in \{M, \ldots, N_{\mathrm{tr}} - 1\}\}$ of local models is obtained. Using $\Gamma$ in an \MMAE{}-like framework would result in an excessively complex scheme. Indeed, \MMAE{} techniques are known to be %
sensitive to the number of employed models, to the point where an excessive number of them can be detrimental to the performance of the overall scheme. To address this issue, we extract a set of $N_{\theta}$ representative models, each described by a parameter vector $\boldsymbol{\gamma}_j, j \in \{1, \ldots, N_{\theta}\}$, using the clustering technique %
in~\cite{masti2021machine}. %
Then, we set $\Gamma^{\mathrm{\SOC{}}} = \{\boldsymbol{\gamma}_j: j \in \{1, \ldots, N_{\theta}\}\}$ and convert each local \ARX{} model %
into its corresponding  minimal state-space representation $\mathcal{M}_j$ in observer canonical form~\cite{mellodge2015practical}. %

\textit{Design of the observer bank~\ref{item:MMAE_2}.}
For each model $\mathcal{M}_j$, we design an observer that provides an estimate $\hat{\boldsymbol{\chi}}_j[k]$ of its state $\boldsymbol{\chi}_j[k]$.
As it will be necessary to run all $N_\theta$ observers in parallel, %
a viable option is to use computationally-light Luenberger observers:
\begin{equation}
  \label{eq:observerDynamic}
  \!\!\!\!\begin{cases}
    \hat{\boldsymbol{\chi}}_j[k\!+\!1]\! = \!A_j \hat{\boldsymbol{\chi}}_j[k] \! + \! B_j i[k] \! + \! d_j \! - \! L_j (\hat{v}_j[k] \! - \! v[k]), \\
    \hat{v}_j[k] = C_j \hat{\boldsymbol{\chi}}_j[k] \!+\! e_j,
  \end{cases}
\end{equation}

where $L_j$ is the observer gain. %
Since minimal state-space realizations are used to define $\mathcal{M}_j$, each pair $(A_j, C_j)$ is fully observable, and the eigenvalues of $A_j-L_jC_j$ can be arbitrarily placed inside the unit circle. %

\textit{Model-free hypothesis testing algorithm~\ref{item:MMAE_3}.}
After the $N_\theta$ observers have been synthesized, we need to build a predictor that replaces the hypothesis testing scheme otherwise used in standard model-based \MMAE{} schemes. To this end, we first extract features from the designed observers~\cite{masti2021machine}. In this work, we employ the absolute values of the innovations $\epsilon_j[k] = \hat{v}_j[k] - v[k], j \in \{1, \ldots, N_{\theta}\},$ as features. The $\epsilon_j[k]$'s are readily obtained by running the observers in~\eqref{eq:observerDynamic} on $\mathcal{D}_{\mathrm{tr}}$ in~\eqref{eq:training_dataset} used to train $M_{\mathrm{\LPV{}}}$. Now, let $\boldsymbol{\varepsilon}[k] \in \mathbb{R}^{N_{\theta} (\ell + 1)}$ be the extracted features vector defined as:

\begin{equation*}
  \boldsymbol{\varepsilon}[k] = \left[
    |\epsilon_1[k]|, \ldots, |\epsilon_1[k\!-\!\ell]|, \ldots, |\epsilon_{N_{\theta}}[k]|, \ldots, |\epsilon_{N_{\theta}}[k\!-\!\ell]|
  \right]^\top,
\end{equation*} 
where $\ell \in \mathbb{N}$ is a window size to be calibrated. Next, we build the augmented training dataset:
\begin{equation}
  \label{eq:augmented_dataset}
  \mathcal{D}_{\mathrm{tr}}^{\mathrm{aug}} \!=\! \{\left(\boldsymbol{\varepsilon}[k], v[k], i[k], \mathrm{\SOC{}}[k]\right)\!: \!k \in \{ \ell, \ldots, N_{\mathrm{tr}}\!- \! 1\}\}\!
\end{equation}
and use it to train a predictor $h_\theta: \mathbb{R}^{N_{\theta} (\ell + 1)} \times \mathbb{R} \times \mathbb{R} \to \mathbb{R}$ such that $h_\theta(\boldsymbol{\varepsilon}[k], i[k], v[k])$ is a good estimate of $\mathrm{\SOC{}}[k]$. This amounts to a standard regression problem:
\begin{equation}
  \label{eq:h_theta_learning}
  \min{h_\theta} \sum_{k = \ell}^{N_{\mathrm{tr}} - 1} \mathcal{L}(h_\theta(\boldsymbol{\varepsilon}[k], i[k], v[k]), \mathrm{\SOC{}}[k]),
\end{equation}
where $\mathcal{L}: \mathbb{R} \times \mathbb{R} \to \mathbb{R}$ is a suitable loss function. Similarly to~\eqref{eq:MLPV_learning}, \ANN{}s can be used as $h_\theta$ and the squared error loss is a common choice for $\mathcal{L}$.

\section{Virtual sensor fusion approach}
\label{sec:sensor_fusion_approach}
As previously pointed out, the \ECM{} in~\eqref{eq:ECM_discrete_time} is only an approximation of true \Liion{} cell behavior compared to~\eqref{eq:general_cell_model}. Instead, the approach reviewed in Section~\ref{sec:data_driven_approach} foregoes the definition of a model but, as shown in~\cite{masti2021machine}, can lead to non-smooth \SOC{} estimates. In this work, we propose to fuse %
an \ECM{}-based Kalman filter %
with the just reviewed machine-learning technique, potentially retaining the smoothness of the \KF{} approach but with the improved accuracy of the data-driven method. To that end, the \ECM{} in~\eqref{eq:ECM_discrete_time} is augmented with an additional output representing the state of charge. Further, the process and measurement noises are added to the state and output equations. In particular, the noises are assumed to be mutually uncorrelated zero-mean white Gaussian noise processes defined as:

\begin{align}
    \label{eq:noises}
    \begin{split}
        & \xi_{\mathrm{\SOC{}}}[k]   \simiid \mathcal{N}\left(0, \sigma_{\mathrm{\SOC{}}}^2\right), \quad   
        \xi_{i_{R}}[k]             \simiid \mathcal{N}\left(0, \sigma_{i_{R}}^2\right),              \\
        &\zeta_{v}[k]               \simiid \mathcal{N}\left(0, \sigma_{v}^2\right), \quad                 
        \zeta_{\mathrm{\SOC{}}}[k] \simiid \mathcal{N}\left(0, \sigma_{\mathrm{\SOC{}}, y}^2\right),
    \end{split}
\end{align}
where $\mathcal{N}\left(\mu, \sigma^2\right)$ is the Gaussian distribution with mean $\mu \in \mathbb{R}$ and variance $\sigma^2 \in \mathbb{R}_{> 0}$. Then, the augmented \ECM{} reads as:
\begin{equation}
    \label{eq:ECM_discrete_time_augmented}
    \begin{cases}
        \mathrm{\SOC{}}\left[k+1\right] = \mathrm{\SOC{}}\left[k\right] - \frac{\tau_{\mathrm{s}}}{Q} \eta\left[k\right] i\left[k\right] + \xi_{\mathrm{\SOC{}}}[k],              \\
        {i}_{R_{1}}\left[k+1\right]     = \exp{-\frac{\tau_{\mathrm{s}}}{\tau_1\left(\mathrm{\SOC{}}[k]\right)}} i_{R_{1}}\left[k\right] +                                        \\
        \qquad + \left(1 - \exp{-\frac{\tau_{\mathrm{s}}}{\tau_1\left(\mathrm{\SOC{}}[k]\right)}}\right) i\left[k\right] + \xi_{i_{R}}[k],                                        \\
        v\left[k\right]                 = \mathrm{\OCV{}} \left(\mathrm{\SOC{}}\left[k\right]\right) - R_{1}\left(\mathrm{\SOC{}}\left[k\right]\right)  i_{R_{1}}\left[k\right] + \\
        \qquad - R_{0}\left(\mathrm{\SOC{}}\left[k\right]\right) i\left[k\right] + \zeta_{v}[k],                                                                                  \\
        \mathrm{\SOC{}}_y\left[k\right] = \mathrm{\SOC{}}\left[k\right] +  \zeta_{\mathrm{\SOC{}}}[k].
    \end{cases}
\end{equation}
We propose to employ the model in~\eqref{eq:ECM_discrete_time_augmented} in a predictor-corrector extended Kalman filter scheme~\cite[Chapter 3]{plett2015battery_volII}. We denote the states of~\eqref{eq:ECM_discrete_time_augmented} estimated by the \EKF{} as 

$\hat{\boldsymbol{x}}^{+}\![k] = [\hat{\mathrm{\SOC{}}}^{+}\!\!\left[k\right], \hat{i}_{R_{1}}^{+}\!\!\left[k\right]]^\top \in \mathbb{R}^2$.
$\hat{\boldsymbol{x}}^{+}\![k]$ is updated based on the input $i[k]$, the terminal voltage measurement and the \SOC{} predicted by the virtual sensor in Section~\ref{sec:data_driven_approach}, which are grouped inside the output vector %
$\boldsymbol{y}[k] = \left[v[k], h_\theta(\boldsymbol{\varepsilon}[k], i[k], v[k]) \right]^\top$. Differently from the baseline \EKF{} that relies only on the measures of $ v[k]$, the proposed observer also considers the predictions $h_\theta(\boldsymbol{\varepsilon}[k], i[k], v[k])$ as \SOC{} measurements and uses them in the computation of the innovations, updating the filter gain accordingly.
Finally, the \EKF{} is initialized with an initial state vector $\hat{\boldsymbol{x}}^{+}\![0] \in \mathbb{R}^2$ and state error covariance matrix $\Sigma_{\tilde{\boldsymbol{x}}}[0] \in \mathbb{R}^{2 \times 2}$ supplied by the user.

\textit{Kalman filter calibration.}
The noise variances in~\eqref{eq:noises} are tuning parameters for the extended Kalman filter that greatly affect its \SOC{} estimation accuracy and smoothness (see, e.g.,~\cite{masti2021machine}). In this work, we propose to calibrate $\sigma_{\mathrm{\SOC{}}}$, $\sigma_{i_{R}}$, $\sigma_{v}$, and $\sigma_{\mathrm{\SOC{}}, y}$, by solving an optimization problem. In particular, consider the dataset $\mathcal{D}_{\mathrm{tr}}$ in~\eqref{eq:training_dataset} previously used for training the data-driven approach in Section~\ref{sec:data_driven_approach}. Given a set of parameters %
$\boldsymbol{\theta}_{\mathrm{\KF{}}} = \left[ \sigma_{\mathrm{\SOC{}}}, \sigma_{i_{R}}, \sigma_{v}, \sigma_{\mathrm{\SOC{}}, y} \right]^\top \in \mathbb{R}^{4}_{>0}$ and an initialization  $\hat{\boldsymbol{x}}^{+}\![0]$, $\Sigma_{\tilde{\boldsymbol{x}}}[0]$, we run the \EKF{} on the data in~\eqref{eq:training_dataset}, obtaining the predicted terminal voltages $\hat{v}[k; \boldsymbol{\theta}_{\mathrm{\KF{}}}]$ and state of charge estimates $\hat{\mathrm{\SOC{}}}^{+}\!\!\left[k; \boldsymbol{\theta}_{\mathrm{\KF{}}}\right]$ for $k \in \{0, \ldots, N_{\mathrm{tr}}-1\}$. The terminal voltage and \SOC{} estimation accuracies are quantified by the Root Mean Squared Errors (\RMSE{}s):

\begin{subequations}
    \label{eq:RMSEs}
    \begin{align}
        J_1(\boldsymbol{\theta}_{\mathrm{\KF{}}})                 & = \sqrt{\frac{1}{N_{\mathrm{tr}}} \sum_{k=0}^{N_{\mathrm{tr}} - 1} (v[k] - \hat{v}[k; \boldsymbol{\theta}_{\mathrm{\KF{}}}])^2},                                              \\
        J_2(\boldsymbol{\theta}_{\mathrm{\KF{}}}) & = \sqrt{\frac{1}{N_{\mathrm{tr}}} \sum_{k=0}^{N_{\mathrm{tr}} - 1} (\mathrm{\SOC{}}[k] - \hat{\mathrm{\SOC{}}}^{+}\!\!\left[k; \boldsymbol{\theta}_{\mathrm{\KF{}}}\right])^2}.
    \end{align}
\end{subequations}
Instead, we employ the Total Variation (\TV{}) as an indicator of \SOC{} estimation smoothness:
\begin{equation}
    \label{eq:SOC_TV}
    \!\!J_3(\boldsymbol{\theta}_{\mathrm{\KF{}}}) \!= \!\frac{\sum_{k=1}^{N_{\mathrm{tr}} \!- \!1}\!\! \left| \hat{\mathrm{\SOC{}}}^{+}\!\!\left[k; \boldsymbol{\theta}_{\mathrm{\KF{}}}\right] \! - \!\hat{\mathrm{\SOC{}}}^{+}\!\!\left[k\!-\!1; \boldsymbol{\theta}_{\mathrm{\KF{}}}\right]\right|}{N_{\mathrm{tr}} \! - \!1}.
\end{equation}
The quantities in~\eqref{eq:RMSEs} and~\eqref{eq:SOC_TV} are combined to give rise to the cost function
\begin{equation}
    \label{eq:cost_fcn_KF}
    J(\boldsymbol{\theta}_{\mathrm{\KF{}}}) = w_1 \frac{J_1(\boldsymbol{\theta}_{\mathrm{\KF{}}})}{v_{\max{}} - v_{\min{}}} + w_2 J_2(\boldsymbol{\theta}_{\mathrm{\KF{}}}) + w_3 J_3(\boldsymbol{\theta}_{\mathrm{\KF{}}}),
\end{equation}
where 
$J_1(\boldsymbol{\theta}_{\mathrm{\KF{}}})$ is normalized to make it assume values that are roughly between $0$ and $1$ (similar to $J_2(\boldsymbol{\theta}_{\mathrm{\KF{}}})$ and $J_3(\boldsymbol{\theta}_{\mathrm{\KF{}}})$), 
while $w_1, w_2, w_3 \in \mathbb{R}_{\geq 0}$ are weights that determine the importance of each term. Finally, the noise variances %
are calibrated by solving: %
\begin{align}
    \label{eq:KF_calibration}
                        & \argmin{\boldsymbol{\theta}_{\mathrm{\KF{}}}} J(\boldsymbol{\theta}_{\mathrm{\KF{}}})                                                                         \\
    \mathrm{s.t.} \quad & \boldsymbol{\theta}_{\mathrm{\KF{}}, \mathrm{lb}} \leq \boldsymbol{\theta}_{\mathrm{\KF{}}} \leq \boldsymbol{\theta}_{\mathrm{\KF{}}, \mathrm{ub}}, \nonumber
\end{align}
where $\boldsymbol{\theta}_{\mathrm{\KF{}}, \mathrm{lb}}, \boldsymbol{\theta}_{\mathrm{\KF{}}, \mathrm{ub}} \in \mathbb{R}^{4}_{>0}$ are user-defined bounds on the parameters. The goal of~\eqref{eq:KF_calibration} is the determination of a \KF{} tuning that leads to \SOC{} estimates that are both accurate and smooth, putting more emphasis on either of the two specifications based on the choice of the weights in~\eqref{eq:cost_fcn_KF}. Given that the computation of the cost in~\eqref{eq:cost_fcn_KF} can be quite time-consuming, requiring running both the \EKF{} and the data-driven virtual sensor %
on the whole $\mathcal{D}_{\mathrm{tr}}$ in~\eqref{eq:training_dataset}, we %
resort to black-box optimization methods~\cite[Chapter 2]{previtali2024phdthesis} to mitigate the time required to solve~\eqref{eq:KF_calibration}.

\section{Experimental results}
\label{sec:experimental_results}
This Section assesses the \SOC{} estimation performance of the proposed Virtual Sensor Fusion (\VSF{}) approach in Section~\ref{sec:sensor_fusion_approach}. %
The method is compared to a Baseline \EKF{} (\BEKF{}), i.e. with no additional output $\mathrm{\SOC{}}_y\left[k\right]$ in~\eqref{eq:ECM_discrete_time_augmented}, and the Virtual Sensor (\VS{}) in Section~\ref{sec:data_driven_approach} on its own. All methods are executed in MATLAB on a machine with an Intel i9-13900H @2.60 GHz CPU and 64 GB of RAM.

\textit{\VS{} training and \EKF{} calibration.}
Consistently with~\cite{masti2021machine}, we employ \ANN{}s, specifically Feed Forward Neural Networks (\FFNN{}s), for $M_{\mathrm{\LPV{}}}$ in~\eqref{eq:MLPV_learning} and $h_\theta$ in~\eqref{eq:h_theta_learning}. In particular, the order $M$ of the \ARX{} model for~\eqref{eq:MLPV_learning} is set to $M = 4$ while $M_{\mathrm{\LPV{}}}$ is an \FFNN{} with two ReLU layers, each with $50$ hidden units, and a linear output layer with $2M+1$ units; the loss $\mathcal{L}_{M_{\mathrm{\LPV{}}}}$ is the absolute error. In line with the results in~\cite{masti2021machine}, the number of representative models is set to $N_{\theta} = 4$ while the gains $L_j$, $j \in \{1, \ldots, N_{\theta}\}$, of the Luenberger observers in~\eqref{eq:observerDynamic} are tuned via pole placement so that all the poles are in the same location, namely around a circle in the complex plane with radius $0.65$. The window length is set to $\ell = 5$. $h_\theta$ in~\eqref{eq:h_theta_learning} is an \FFNN{} with two ReLU layers, each with $30$ hidden units, and a linear output layer with $1$ unit; the loss $\mathcal{L}$ is the squared error. The network architectures of $M_{\mathrm{\LPV{}}}$ and $h_\theta$, the \ARX{} order\footnote{The bias term of the affine model in~\eqref{eq:MLPV_learning} was discarded after training due to it being negligible.} $M$, and the window length $\ell$ were chosen in such a way as to achieve the best results on a portion ($20\%$) of $\mathcal{D}_{\mathrm{tr}}$ in~\eqref{eq:training_dataset} reserved for validation purposes. The validation dataset is also used for early stopping during \ANN{} training. Neural network training was carried out via MATLAB's Deep Learning Toolbox using the RMSprop optimizer, running for $100$ epochs.

The \EKF{} is initialized as %
$\hat{\boldsymbol{x}}^{+}[k] = \left[0, 0\right]^\top$ (cell at rest and fully discharged) and $\Sigma_{\tilde{\boldsymbol{x}}}[0] = \diag{0.5, 0.001}$ (high initial \SOC{} uncertainty). Problem~\eqref{eq:KF_calibration} is solved by means of the \texttt{GLIS-r} \BBO{} algorithm~\cite[Chapter 5]{previtali2024phdthesis} with bounds $\boldsymbol{\theta}_{\mathrm{\KF{}}, \mathrm{lb}} = 10^{-6} \cdot \boldsymbol{1}_4$ and $\boldsymbol{\theta}_{\mathrm{\KF{}}, \mathrm{ub}} = \boldsymbol{1}_4$, $\boldsymbol{1}_4$ being the $4$-dimensional column vector of ones. The weights for~\eqref{eq:cost_fcn_KF} are set to $w_1 = 0.5, w_2 = 1, w_3 = 5$ to favor \SOC{} estimation accuracy over terminal voltage \RMSE{} while giving great emphasis to smoothness. To ensure a fair comparison, the \BEKF{} is tuned %
in a similar fashion.

\textit{Results.}
\begin{figure*}[!htb]
	\centering
    \includegraphics[width=\textwidth]{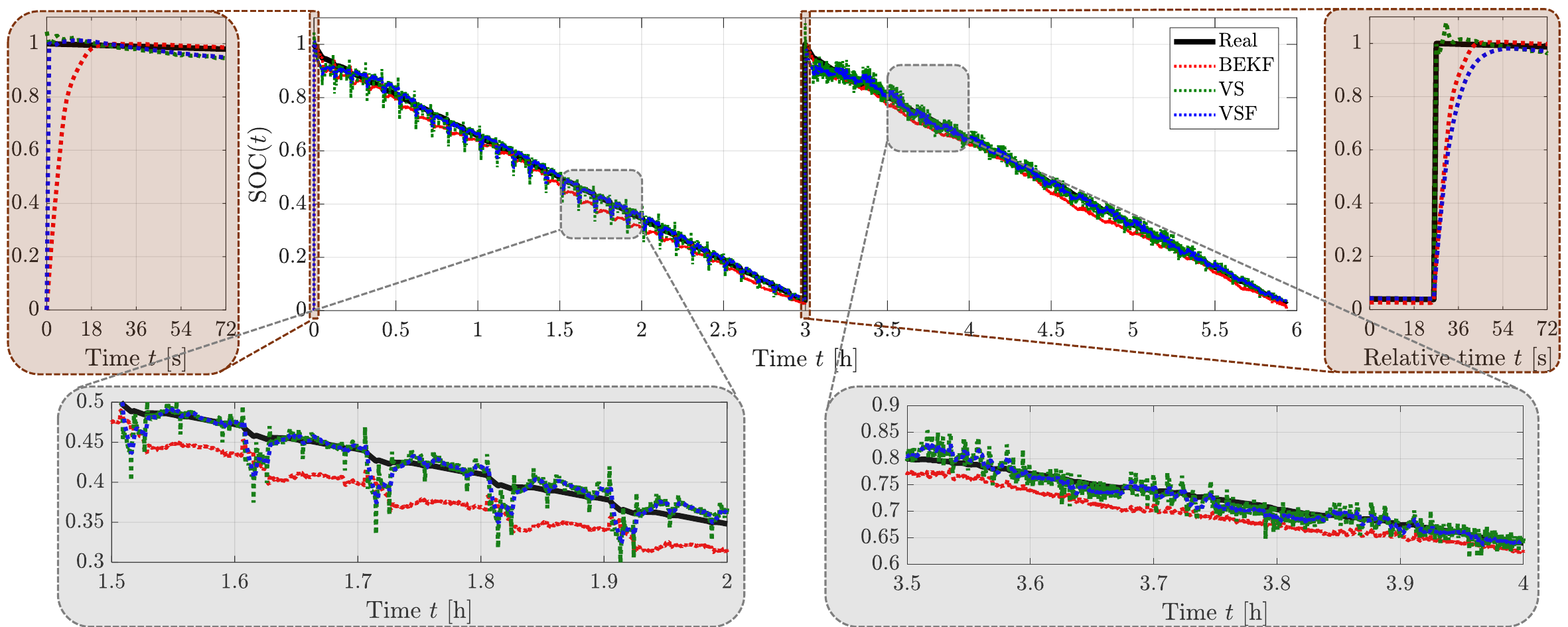}
	\caption{\label{fig:SOC_estimation results} State of charge estimated by the considered methods on the test dataset $\mathcal{D}_{\mathrm{tst}}$ in~\eqref{eq:test_dataset} (DST + US06).}
\end{figure*}
The performances of the methods under study are assessed on the test dataset $\mathcal{D}_{\mathrm{tst}}$ in~\eqref{eq:test_dataset}, which was neither used for \VS{} training nor \KF{} calibration. The results are presented in \figurename{}~\ref{fig:SOC_estimation results}. We can see that the \SOC{} estimated by the \BEKF{}, although very smooth, exhibits a non-negligible underestimation error that is consistent throughout the experiment, amounting to roughly $2.2\%$ median-wise. Instead, the \VS{} closely follows the real \SOC{} but much less smoothly than the \BEKF{}. Notably, the proposed \VSF{} approach compensates the \SOC{} underestimation issue while achieving a level of smoothness that is in between the two methods. %
In any case, all methods can handle the abrupt \SOC{} transitions quite well, converging (roughly) to the true \SOC{} in a negligible amount of time.

For further analysis, in \tablename{}~\ref{tab:performance}, we report the \SOC{} \RMSE{}s in~\eqref{eq:RMSEs} and \TV{}s in~\eqref{eq:SOC_TV} for both datasets in~\eqref{eq:dynamic_current_profiles_data} along with the median computational times for a single step of each procedure. We can notice that the \BEKF{} is the least accurate (highest \RMSE{}s) but the computationally-lightest. Instead, the \VS{} is the most accurate but also the least smooth (highest \TV{}s). Notably, the \VSF{} approach achieves \RMSE{}s that are roughly $19\%$ higher than the \VS{} but \TV{}s that are more than $80\%$ lower than the virtual sensor on its own. %
Consequently, we can say that the proposed method greatly improves \SOC{} estimation smoothness compared to the \VS{} at the cost of a slight decrease in accuracy. Surprisingly, the \VSF{} technique also outperforms the \BEKF{} in terms of \TV{}. Finally, the \VS{} and \VSF{} methods are the most time-consuming but their computational overhead is practically negligible w.r.t. the employed sampling time $\tau_{\mathrm{s}} = 1\,\mathrm{s}$ (Section~\ref{sec:experimental_setup}).
\begin{table}[!htb]
    \centering
    \caption{\SOC{} estimation performance and computational times of the methods. Best results are highlighted with a bold font.}
    \settablestretch
    \settablefontsize
    \label{tab:performance}
    
    \begin{tabular}{cccccc}
        & \multicolumn{2}{c}{\textbf{BJDST+FUDS}} & \multicolumn{2}{c}{\textbf{DST+US06}} & \tabularnewline
        & $\mathrm{RMSE}$ & $\mathrm{TV}$ & $\mathrm{RMSE}$ & $\mathrm{TV}$ & Time $\left[\mathrm{s}\right]$\tabularnewline
       \hline 
       \colorrowoftable{}\textbf{\BEKF{}} & $0.0315$ & $0.0011$ & $0.0308$ & $0.0013$ & $\boldsymbol{6.2 \cdot 10^{-5}}$\tabularnewline
       \VS{} & $\boldsymbol{0.0161}$ & $0.0048$ & $\boldsymbol{0.0159}$ & $0.0060$ & $9.5 \cdot 10^{-4}$\tabularnewline
       \colorrowoftable{}\textbf{\VSF{}} & $0.0192$ & $\boldsymbol{0.0009}$ & $0.0188$ & $\boldsymbol{0.0009}$ & $1.1 \cdot 10^{-3}$\tabularnewline
       \hline 
       \end{tabular}
\end{table}

\section{Conclusion}
\label{sec:conclusion}
This paper investigates the combination of Kalman filters based on equivalent-circuit models with %
machine-learning methods for state of charge estimation of \Liion{} cells. Particularly, the method in~\cite{masti2021machine} is considered, which amounts to a virtual sensor that relies on a bank of observers extracted from an \APV{} \ARX{} model learnt directly from data followed by an \FFNN{} for \SOC{} prediction. %
The predictions of the virtual sensor are fed to an \EKF{} that employs an augmented \ECM{} for smoothing purposes. Its noise covariance matrices are optimized to achieve a good trade-off between \SOC{} estimation accuracy and smoothness. Experimental results show that the proposed %
approach (i) outperforms the baseline \EKF{} on both specifications, (ii) attains slightly higher \RMSE{}s ($+19\%$) but greatly lower \TV{}s ($-80\%$) compared to the virtual sensor on its own, (iii) exhibits a negligible computational cost.
Future work is devoted to investigating the effect of temperature and aging effects on the \SOC{} estimation accuracy of our proposal.

\section*{Acknowledgments}
This study was carried out within the MOST - Sustainable Mobility National Research Center and received funding from the European Union NextGenerationEU (PIANO NAZIONALE DI RIPRESA E RESILIENZA (PNRR) - MISSIONE 4 COMPONENTE 2, INVESTIMENTO 1.4 - D.D. 1033 17/06/2022, CN00000023), Spoke 5 \quotes{Light Vehicle and Active Mobility}. This manuscript reflects only the authors' views and opinions, neither the European Union nor the European Commission can be considered responsible for them.
This work has been partially funded by the European Union - NextGenerationEU under the Italian Ministry of University and Research (MUR) National Innovation Ecosystem grant ECS00000041 - VITALITY – CUP: D13C21000430001. Daniele Masti is also part of INdAM/GNAMPA.

\bibliographystyle{plain}
{
  \bibliography{ref.bib}
}

\end{document}